\documentclass[conference]{IEEEtran}
\usepackage[latin1,utf8]{inputenc}

\usepackage[cyr]{aeguill}

\usepackage[
  pass,
]{geometry}

\usepackage{amsmath,bbm,comment,subfigure,balance,flushend}

\usepackage[dvipsnames]{xcolor}
\usepackage{listings}

\usepackage{graphicx}

\usepackage{lipsum}

\usepackage{acronym}
\usepackage[acronym]{glossaries}
\usepackage{glossary-mcols}
\usepackage{url}
\usepackage{booktabs}
\def\tallafigura{0.29} 
\def\tallafigurabig{0.4} 
\def\tallafigurabigg{0.35} 

\usepackage{amsmath}
\usepackage{algorithm}
\usepackage[noend]{algpseudocode}
\usepackage{siunitx} 

\usepackage{array}
\usepackage[T1]{fontenc}

\makeatletter
\def\BState{\State\hskip-\ALG@thistlm}
\makeatother

\def\P{\mathbbm{P}}

\def\E{\mathbbm{E}}

\newacronym{AAA}{AAA}{Authentication, Authorization, and Accounting}
\newacronym{RSC}{RSC}{Recursive  Systematic Convolutional}
\newacronym{LLR}{LLR}{Log-Likelihood Ratio}
\newacronym{FS}{FS}{Functional Split}
\newacronym{BBU}{BBU}{Base Band Unit}
\newacronym{COTS}{COTS}{Commercial off-the-shelf }
\newacronym{VNF}{VNF}{Virtualized Network Function}
\newacronym{VNF FG}{VNF FG}{VNF Forwarding Graph}
\newacronym{NFV}{NFV}{Network Function Virtualization}
\newacronym{GPP}{GPP}{General Purpose Processor}
\newacronym{vEPC}{vEPC}{virtual Evolved Packet Core}
\newacronym{LTE}{LTE}{Long Term Evolution}
\newacronym{uRLLC}{uRLLC}{Ultra-Reliable Low-Latency Communications}
\newacronym{eMBB}{eMBB}{enhanced Mobile BroadBand}
\newacronym{mMTC}{mMTC}{massive Machine Type Communications}
\newacronym{OSM}{OSM}{Open Source MANO}
\newacronym{C-EPC}{C-EPC}{Cloud-EPC}
\newacronym{EPCaaS}{EPCaaS}{EPC as a Service}
\newacronym{TDD}{TDD}{Time Division Duplex}
\newacronym{UE}{UE}{User Equipment}
\newacronym{HARQ}{HARQ}{Hybrid Automatic Repeat-Request}
\newacronym{PRB}{PRB}{Physical Resource Blocks}
\newacronym{MCS}{MCS}{Modulation and Coding Scheme}
\newacronym{CQI}{CQI}{Channel Quality Indicator}
\newacronym{DC}{DC}{Dedicated Core}
\newacronym{RR}{RR}{Round Robin}
\newacronym{G}{G}{Greedy}
\newacronym{VPN}{VPN}{Virtual Private Network}
\newacronym{MPLS}{MPLS}{Multiprotocol Label Switching}
\newacronym{OWL}{OWL}{Web Ontology Language}
\newacronym{NST}{NST}{Network Slice Template}
\newacronym{NSST}{NSST}{Network Slice Subnet Template}
\newacronym{NSMF}{NSMF}{Network Slice Management Function}
\newacronym{NSSMF}{NSSMF}{Network Slice Subnet Management Function}
\newacronym{CSMF}{CSMF}{Communication Service Management Function}
\newacronym{FCAPS}{FCAPS}{Fault-management Configuration Accounting Performance and Security}

\newacronym{CNF}{CNF}{Cloud-Native Network Function}
\newacronym{PLMN}{PLMN}{Public Land Mobile Network}
\newacronym{CLAMP}{CLAMP}{Closed Loop Automation Management Platform}

\newacronym{FDD}{FDD}{Frequency Division Duplex}
\newacronym{OFDM}{OFDM}{Orthogonal Frequency Division Multiplexing}

\newacronym{VM}{VM}{Virtual Machine}
\newacronym{PDCP}{PDCP}{Packet Data Convergence Protocol}
\newacronym{MAC}{MAC}{Medium Access Control}
\newacronym{RLC}{RLC}{Radio Link Control}
\newacronym{RRC}{RRC}{Radio Resource Control}
\newacronym{AM}{AM}{Acknowledged Mode}
\newacronym{UM}{UM}{Unacknowledged Mode}
\newacronym{TM}{TM}{Transparent Mode}
\newacronym{MIMO}{MIMO}{Multiple Input Multiple Output}
\newacronym{MISO}{MISO}{Multiple Input Single Output}
\newacronym{SIMO}{SIMO}{Single Input Multiple Output}
\newacronym{SISO}{SISO}{Single Input Single Output}
\newacronym{MCC}{MCC}{Mobile Country Code}
\newacronym{MNC}{MNC}{Mobile Network Code}
\newacronym{S-TMSI}{S-TMSI}{Shortened Temporary Mobile Subscriber Identity}
\newacronym{IMSI}{IMSI}{International Mobile Subscriber Identity}
\newacronym{DRB}{DRB}{Dedicated Radio Bearer}
\newacronym{GUMMEI}{GUMMEI}{Globally Unique MME Identity}

\newacronym{PCI}{PCI}{Physical-layer Cell Identity}
\newacronym{ROHC}{ROHC}{Robust Header Compression}
\newacronym{SN}{SN}{Sequence Number}
\newacronym{RAR}{RAR}{Random Access Response}
\newacronym{C-RNTI}{C-RNTI}{Cell Radio Network Temporary Identifier}
\newacronym{BSR}{BSR}{Buffer Status Report}
\newacronym{DRX}{DRX}{Discontinuous Reception}
\newacronym{PHR}{PHR}{Power Head Room}
\newacronym{PUSCH}{PUSCH}{Physical Uplink Shared Channel}
\newacronym{ADM}{ADM}{Activation/Deactivation MAC}
\newacronym{GP}{GP}{Gap Period}

\newacronym{RE}{RE}{Resource Element}
\newacronym{RB}{RB}{Resource Block}
\newacronym{REG}{REG}{Resource Element Group}
\newacronym{CSRS}{CSRS}{Cell-Specific Reference Signal}
\newacronym{IFFT}{IFFT}{Inverse Fast Fourier Transform}
\newacronym{OFDMA}{OFDMA}{Orthogonal Frequency Division Multimple Access}
\newacronym{CRC}{CRC}{Cyclic Redundancy Check}
\newacronym{SFC}{SFC}{Service Function Chain}

\newacronym{eNB}{eNB}{Evolved NodeB}
\newacronym{RAN}{RAN}{Radio Access Network}
\newacronym{ARQ}{ARQ}{Automatic Repeat reQuest}
\newacronym{NAS}{NAS}{Non-Access Stratum}
\newacronym{MME}{MME}{Mobility Management Entity}
\newacronym{MIB}{MIB}{Master Information Block}
\newacronym{SIB}{SIB}{System Information Block}
\newacronym{RSRP}{RSRP}{Reference Signal Received Power}
\newacronym{RAT}{RAT}{Radio Access Technologie}
\newacronym{ACK}{ACK}{Acknowledge}
\newacronym{NACK}{NACK}{Negative acknowledge}
\newacronym{PDCCH}{PDCCH}{Physical Downlink Control Channel}
\newacronym{SAW}{SAW}{Stop and Wait}
\newacronym{TTI}{TTI}{Transmission Time Interval}
\newacronym{RRH}{RRH}{Radio Remote Head}
\newacronym{SNIR}{SNIR}{Signal-to-Noise-plus-Interference Ratio}
\newacronym{WCET}{WCET}{Worst Case Execution Time}
\newacronym{GPC}{GPC}{General Purpose Computer}
\newacronym{KPI}{KPI}{Key Performance Indicator}
\newacronym{OAI}{OAI}{Open Air Interface}
\newacronym{IMS}{IMS}{IP Multimedia Subsystem}
\newacronym{vIMS}{vIMS}{virtual IP Multimedia Subsystem}
\newacronym{EPC}{EPC}{Evolved Packet Core}
\newacronym{SDN}{SDN}{Software Defined Network}
\newacronym{C-RAN}{C-RAN}{Centralized-RAN}
\newacronym{OS}{OS}{Operating System}
\newacronym{TB}{TB}{Transport Block}
\newacronym{TBS}{TBS}{Transport Block Size}
\newacronym{QCI}{QCI}{QoS Channel Indicator}

\newacronym{GPU}{GPU}{Graphics Processing Unit}
\newacronym{CPU}{CPU}{Central Processing Unit}

\newacronym{SDU}{SDU}{Service Data Unit}
\newacronym{CBS}{CBS}{Code Block Size}
\newacronym{CB}{CB}{Code Block}
\newacronym{SPMD}{SPMD}{Single Program Multiple Data}
\newacronym{SIMD}{SIMD}{Single Instruction Multiple Data} 

\newacronym{SINR}{SINR}{Signal-to Interference Noise Ratio}
\newacronym{CO}{CO}{Central Office}
\newacronym{CA}{CA}{Carrier Aggregation}
\newacronym{SRS}{SRS}{Sound Reference Signal}
\newacronym{SC-OFDMA}{SC-OFDMA}{Single Carrier - Orthogonal Frequency Division Multiple Access}
\newacronym{FPGA}{FPGA}{Field-Programmable Gate Array}
\newacronym{TA}{TA}{Time Advancing}
\newacronym{CoMP}{CoMP}{Coordinated Multi-point}
\newacronym{NPRB}{NPRB}{Number of Physical Resource Blocks}
\newacronym{RTT}{RTT}{Round Trip Time}
\newacronym{CPRI}{CPRI}{Common Public Radio Interface}
\newacronym{CBR}{CBR}{Constant Bit Rate}
\newacronym{NRB}{NRB}{Number of Resource Blocks}
\newacronym{BJF}{BJF}{Biggest Job First}
\newacronym{EDF}{EDF}{Earliest Deadline First}
\newacronym{FCFS}{FCFS}{First-come, First-served}
\newacronym{PSTN}{PSTN}{Public Switched Telephone Network}
\newacronym{ETSI}{ETSI}{European Telecommunications Standards Institute}
\newacronym{vBBU}{vBBU}{virtualized BBU}
\newacronym{vRAN}{vRAN}{virtualized RAN}
\newacronym{IoT}{IoT}{Internet of Things}
\newacronym{B2B}{B2B}{Business to Business}
\newacronym{B2C}{B2C}{Business to Customer}
\newacronym{QoE}{QoE}{Quality of Experience}
\newacronym{QoS}{QoS}{Quality of Service}
\newacronym{VNO}{VNO}{Virtual mobile Network Operator}
\newacronym{SLA}{SLA}{Service Level Agreement}
\newacronym{VRRM}{VRRM}{Virtual Radio Resource Management}
\newacronym{KVM}{KVM}{Kernel-based Virtual Machine}
\newacronym{LXC}{LXC}{Linux Containers}
\newacronym{PS}{PS}{Processor Sharing}

\newacronym{eCPRI}{eCPRI}{evolved CPRI}
\newacronym{RoE}{RoE}{Radio over Ethernet}
\newacronym{PAPR}{PAPR}{Peak-to-average power ratio}
\newacronym{SC-FDMA}{SC-FDMA}{Single Carrier Frequency Division Multiple Access}
\newacronym{AGC}{AGC}{Automatic Gain Control}
\newacronym{PMD}{PMD}{Polarization Mode Dispersion}
\newacronym{ADC}{ADC}{Analogic-Digital Converter}

\newacronym{IQ}{IQ}{In-Phase Quadrature}
\newacronym{xRAN}{xRAN}{extensible Radio Access Network}
\newacronym{ISI}{ISI}{Inter-symbol interference}
\newacronym{FFT}{FFT}{Fast Fourier Transform}
\newacronym{IPC}{IPC}{Inter process communication}
\newacronym{CCDU}{CCDU}{Channel Coding Data Unit}
\newacronym{CC}{CC}{Channel Coding}
\newacronym{gNB}{gNB}{next-Generation Node B}
\newacronym{EUTRAN}{EUTRAN}{Evolved Universal Terrestrial Radio Access Network}
\newacronym{SCTP}{SCTP}{Stream Control Transmission Protocol}
\newacronym{NR}{NR}{New Radio}
\newacronym{NF}{NF}{Network Function}
\newacronym{CU}{CU}{Central Unit}
\newacronym{DU}{DU}{Distributed Unit}
\newacronym{NGC}{NGC}{Next Generation Core}
\newacronym{DL}{DL}{down-link}
\newacronym{UL}{UL}{up-link}
\newacronym{LJF}{LJF}{Largest Job First}
\newacronym{RANaaS}{RANaaS}{RAN as a Service}
\newacronym{NaaS}{NaaS}{Network as a Service}
\newacronym{NS}{NS}{Network Service}
\newacronym{FG}{FG}{Forwarding Graph}
\newacronym{VNFC}{VNFC}{VNF Component}
\newacronym{MANO}{MANO}{Management and Orchestration}
\newacronym{FIFO}{FIFO}{First In Firs Out}
\newacronym{NFVI}{NFVI}{NFV Infrastructure}
\newacronym{NFVO}{NFVO}{NFV Orchestrator}
\newacronym{PoP}{PoP}{Point of Presence}
\newacronym{NAT}{NAT}{Network Address Translation}
\newacronym{CDN}{CDN}{Content Delivery Network}
\newacronym{VNFM}{VNFM}{VNF Manager}
\newacronym{EM}{EM}{Element Management}
\newacronym{VIM}{VIM}{Virtualised Infrastructure Manager}

\newacronym{e2e}{e2e}{end-to-end}

\newacronym{AMF}{AMF}{Access and Mobility Management Function}
\newacronym{SMF}{SMF}{Session Management Function}
\newacronym{UPF}{UPF}{User Plane Function}
\newacronym{PCF}{PCF}{Policy Control Function}
\newacronym{UDM}{UDM}{Unified Data Management}
\newacronym{NRF}{NRF}{NF Repository Function}
\newacronym{AUSF}{AUSF}{Authentication Server Function}
\newacronym{API}{API}{Application Programming Interface}
\newacronym{HSS}{HSS}{Home Subscriber Server}
\newacronym{PCRF}{PCRF}{Policy and Charging Rules Function}
\newacronym{SOA}{SOA}{Software-Oriented Architecture}
\newacronym{AKA}{AKA}{Authentication and Key Agreement}
\newacronym{AF}{AF}{Application Function}
\newacronym{NEF}{NEF}{Network Exposure Function}
\newacronym{NSSF}{NSSF}{Network Slice Selection Function}
\newacronym{NSSP}{NSSP}{Network Slice Service Profile}
\newacronym{VES}{VES}{Virtual Event Streaming}
\newacronym{NSSAI}{NSSAI}{Network Slice Selection Assistance Information}

\newacronym{NSSI}{NSSI}{Network Slice Subnet Instance}

\newacronym{NSS}{NSS}{Network Slice Subnet}
\newacronym{NSC}{NSC}{Network Slice Customer}
\newacronym{NSP}{NSP}{Network Slice Provider}
\newacronym{CSC}{CSC}{Communication Service Customer}
\newacronym{CSP}{CSP}{Communication Service Provider}
\newacronym{SST}{SST}{Slice/Service Type}
\newacronym{SD}{SD}{Slice Differentiator}
\newacronym{USRP}{USRP}{UE Router Selection Policy}
\newacronym{S-NSSAI}{S-NSSAI}{Single Network Slice Selection Assistance Information}
\newacronym{ONISTT}{ONISTT}{Open Net-centric Interoperability Standards for Training and Testing}
\newacronym{KB}{KB}{Knowledge Base}
\newacronym{NSI}{NSI}{Network Slice Instance}

\newacronym{VF}{VF}{Virtual Function}
\newacronym{VFC}{VFC}{Virtual Function Component}
\newacronym{CR}{CR}{Complex Resource}
\newacronym{PNF}{PNF}{Physical Network Function}
\newacronym{CP}{CP}{Connection Point}
\newacronym{VL}{VL}{Virtual Link}
\newacronym{SDC}{SDC}{Service Design and Creation}
\newacronym{ONAP}{ONAP}{Open Network Automation Platform}
\newacronym{VID}{VID}{Virtual Infrastructure Deployment}
\newacronym{VSP}{VSP}{Vendor Software Product}

\newacronym{WEF}{WEF}{Wireless Edge Factory}

\newacronym{DP}{DP}{Data Plane}
\newacronym{ECOMP}{ECOMP}{Enhanced Control Orchestration Management and Policy}

\newacronym{AAI}{AAI}{Active and Available Inventory}
\newacronym{SDNC}{SDNC}{Software Defined Network Controller}
\newacronym{SO}{SO}{Service Orchestrator}
\newacronym{APPC}{APPC}{Application Controller}
\newacronym{DCAE}{DCAE}{Data Collection Analytics and Events}
\newacronym{OOF}{OOF}{ONAP Optimization Framework}
\newacronym{OSS}{OSS}{Operation Support System}
\newacronym{BSS}{BSS}{Business Support System}
\newacronym{SOCKS}{SOCKS}{Secured Over Credential-based Keberos}
\newacronym{VVP}{VVP}{VNF Validation Program}
\newacronym{PDP}{PDP}{Policy Decision Point}
\newacronym{PEP}{PEP}{Policy Enforcement Point}
\newacronym{PCC}{PCC}{Policy Creation Component}

\newacronym{VLM}{VLM}{Vendor License Model}

\newacronym{CUPS}{CUPS}{Control User Plane Separation}

\newglossarystyle{modsuper}{%
  \glossarystyle{super}%

}

\begin{document}

\title{Contribution to the design and the implementation of a Cloud Radio Access Network}

\author{\IEEEauthorblockN{Veronica Quintuna Rodriguez\IEEEauthorrefmark{1}, Fabrice Guillemin\IEEEauthorrefmark{1} and Philippe Robert\IEEEauthorrefmark{2}}\\ 
\IEEEauthorblockA{\IEEEauthorrefmark{1}Orange Labs, 2 Avenue Pierre Marzin,  22300 Lannion, France} \IEEEauthorrefmark{2}Inria, 2 rue Simone Iff, 75589 Paris, France\\ \{veronica.quintunarodriguez, fabrice.guillemin\}@orange.com, philippe.robert@inria.fr}

\maketitle  
\begin{abstract}
 This dissertation paper presents the main contributions to the design and the implementation of a Cloud-RAN solution. We concretely address the two main challenges of Cloud-RAN systems: real-time processing of radio signals and reduced fronthaul capacity. We propose a  multi-threading model to achieve latency reduction of critical RAN functions as well as an adapted functional split for optimizing the transmission of radio signals. We model the performance of the proposed solution by means of stochastic service systems which reflect the behavior of high performance computing architectures based on parallel processing and yield dimensioning rules for the required computing capacity. Finally, we validate the accuracy of the theoretical proposals by a Cloud-RAN testbed implemented on the basis of open source solutions, namely Open Air Interface (OAI).  



 
 
\end{abstract}



\section{Introduction}

The emergence of the virtualization technology plays a crucial role in the evolution of telecommunications network architectures, notably by enabling \gls{NFV}. This is clearly a groundbreaking evolution in the design of future networks and IT infrastructures, which will eventually be completely merged. NFV moreover promises   significant economic savings as well as more flexible and accurate management of resources.   \gls{NFV} precisely consists of decoupling network functions from their hosting hardware. Network operators are thus able to instantiate on the fly \glspl{VNF}, which appear as software suites running at various network locations in order to meet customer requirements.

The main goal of this PhD thesis~\cite{inriaMyPhD} is to investigate the  performance of \glspl{VNF} by considering  a driving use case. In this PhD, we have chosen Cloud-RAN, introduced in earlier papers (see for instance \cite{chinaMobile}). Cloud-RAN notably presents ramifications in terms of network design and is emblematic in terms of performance. The design of Cloud-RAN includes (i) the analysis in terms of architecture, (ii) the identification of  \glspl{KPI} that reflect the performance, (iii) the decomposition of global network functions into elementary components, (iv) the dimensioning of cloud resources by taking into account the scheduling strategy and (v) the validation of the  theoretical models by means of a testbed implementation.

Cloud-RAN aims at virtualizing and centralizing higher-\gls{RAN} functions in the network while keeping lower-RAN functions in distributed units (near to antennas). These two nodes, so-called respectively, \gls{CU} and \gls{DU} by the 3GPP enable flexible and scalable functional splits which can be adapted to the required network performance. In addition, the co-location of \glspl{CU} with Mobile/Multi-access Edge Computing facilities opens the door to the realization of low latency services, thus meeting the strict requirements of Ultra Reliable Low Latency Communications (URLLC)~\cite{3GPP38_801}.

The implementation of Cloud-RAN  actually raises many issues, in particular with regard to the network architecture (CU/DU placement, required bandwidth for the fronthaul, etc.) as well as to resources allocation (required computing capacity for processing real-time radio signals in the cloud). The collocation of various virtual \glspl{BBU} in the cloud however enables the possibility of joined radio resource allocation for interference reduction and data rate improvements. Similarly, the centralized base band processing of radio signals in the cloud allows applying resource pooling and statistical multiplexing principles when allocating computing resources. These claims are in the core of this study. 

Furthermore, we notably focus on improving the processing time of virtual RAN functions in the aim of increasing the front-haul time-budget (distance between DU and CU), and as a consequence to reach a higher aggregation of virtual radio cells in the \gls{CO}. In this work, we particularly propose a multi-threading model based on parallel processing to reduce the runtime of encoding/decoding functions on general purpose computers. Beyond low latency processing, we introduce an adapted functional split in order to keep the most resource consuming RAN functions (namely, the channel encoding/decoding function) in the \gls{CU} while optimizing the required fronthaul capacity.

For dimensioning purposes, we specifically introduce a batch queuing model, namely the $M^{[X]}/M/C$ multi-service system, to assess the needed processing capacity in a data center while meeting the RAN latency requirements. The model particularly considers the execution of base band functions under the principle of  high-performance parallel programming on multi core systems, i.e., parallel runnable jobs are executed at the \emph{same} physical instant on separate cores. In addition, we studied the $M^{[X]}/M/1$~$PS$ for modeling the parallel processing in concurrent environments, i.e., when various jobs share a single processing unit by interleaving execution steps of each process via time-sharing slices (also referred to as time-slots)~\cite{pike2012concurrency,silberschatz2014operating}.

In order to confirm the accuracy of the theoretical models, we perform numerical experiments using statistical parameters captured from the Cloud-RAN emulation while using \gls{OAI}, an open-source solution which implements the RAN functionality in software. Beyond simulation, as a proof of concept,  we implement the proposed multi-threading models in a OAI-based test-bed. Performance results show important gains in terms of latency which make true the promises of fully centralized Cloud-RAN architectures and enable the cloudification of critical network functions.

This paper is organized as follows: In Section~\ref{analysis}, we introduce a general analysis of Cloud-RAN and the design guidelines for implementing a Cloud RAN solution. In Section~\ref{design}, we model the behavior of Cloud-RAN by means of stochastic service systems for dimensioning purposes. The theoretical models are validated in Section~\ref{poc}, we notably describe a proof of concept and the main performance results. Concluding remarks and research perspectives are presented in Section~\ref{conclusion}.




%



\section{Cloud-RAN Analysis and Implementation Guidelines}
\label{analysis}

Cloud-RAN, also referred to as C-RAN, aims at implementing the whole base-band processing of radio signals in software while keeping distributed antennas and gathering \glspl{BBU} in a \gls{CO}. A virtual \gls{BBU} implements in software all network functions belonging to the three lower layers of the E-UTRAN protocol stack. These functions mainly concern PHY functions as signal generation, IFFT/FFT, modulation and demodulation, encoding and decoding; radio scheduling; concatenation/segmentation of \gls{RLC} protocol; and encryption/decryption procedures of \gls{PDCP}, for the down-link and up-link directions~\cite{rodriguez2017towards,rodriguez2017vnf}. 
In Cloud-RAN systems, the whole base-band processing of radio signals must meet strict latency requirements (namely, $1$ millisecond in the down-link direction and $2$ milliseconds in the up-link). 

Virtualizing network functions and running them on distant servers raise many issues in terms of performance and network control. The case of Cloud-RAN exemplifies the complexity of virtualizing real-time network functions, however virtualizing and orchestrating an end-to-end mobile network involves additional intricate issues (see for instance~\cite{NoF}).

%

\subsection{Fronthaul analysis}
One of the main issues of Cloud-RAN is the required bandwidth to transmit radio signals between the BBU-pool (namely, \glspl{CU}) and each DU placed near to antennas. When considering current used transmission protocols as \gls{CPRI}, the required fronthaul capacity strictly depends of the number of virtual radio cells hosted in the data center (\gls{CO}) and not of the traffic in the cells. 



The fronthaul capacity problem relies not only on the constant bit rate used by \gls{CPRI}~\cite{duan2016performance} but on the high redundancy present in the transmitted I/Q signals. Many efforts are currently being devoted to reduce optic-fiber resource consumption such as I/Q compression~\cite{guo2013lte}, non-linear quantization, sampling rate reduction, and even, packetization of \gls{CPRI}. Several functional splits of the physical layer are also being an object of study in order to save fiber bandwidth~\cite{wubben2014benefits,duan2016performance}. The required fronthaul capacity for the various functional splits considered by 3GPP is analyzed in~\cite{icin2019}. It turns out that the required fronthaul capacity significantly decreases when the functional split is shifted after the PHY layer or even after the MAC layer~\cite{wubben2014benefits}, however, these architectures do not enable exploiting the advantages of the base band centralization. 

\begin{figure}[hbtp]
\centering
  \includegraphics[scale=0.46, trim=0 260 100 0, clip] {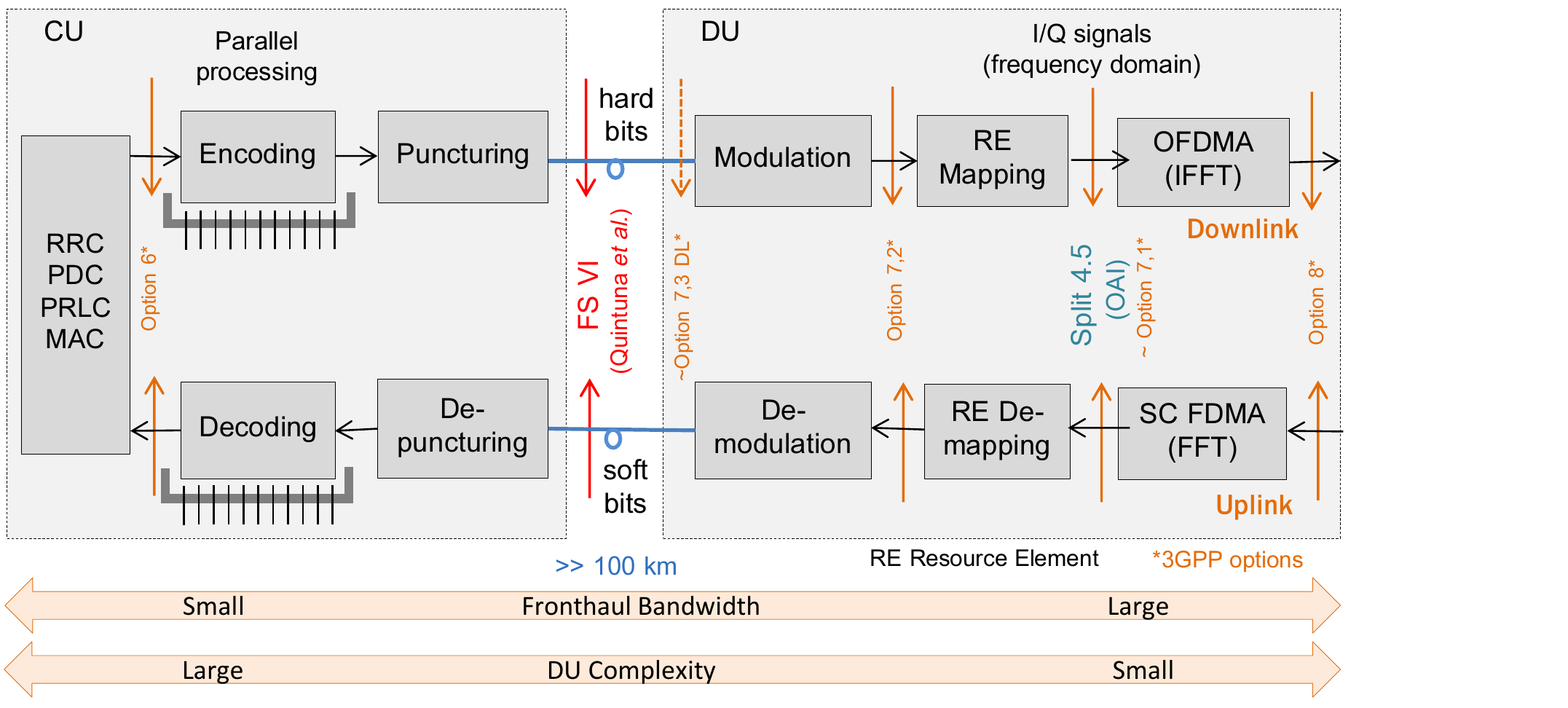}
    \caption{Functional split architecture.}
    \label{figsplit}
\end{figure}

In the following, we focus in a fully centralized RAN architecture that processes the most resource consuming base band functions in the cloud. In view of the analysis carried out in~\cite{inriaMyPhD,icin2019}, we adopt a bidirectional intra-PHY split (referred to as Functional Slit VI). This split centralizes the channel encoding/decoding functions and keeps the modulation/demodulation functions near to antennas. As shown in Figure~\ref{figsplit}, the split transmits hard bits in the downlink and soft bits (real and not binary values) in the uplink. The soft bits represent the \gls{LLR}, i.e., the radio of the probability that a particular bit was 1 and the probability that the same bit was 0. (Log is used for better precision).  The split notably enables a significant gain when comparing it to the initial \gls{CPRI} solution (the required fronthaul capacity in the downlink is then up to $100$ Mbps. when using normal cyclic prefix and the maximum modulation order supported in LTE)~\cite{icin2019}.

The hard/soft bits are encapsulated into Ethernet frames and transmitted by fiber links. Namely, \gls{RoE} is considered by IEEE Next Generation fronthaul Interface (1914) Working Group as well as by the xRAN fronthaul Working Group of the xRAN Forum. The proposed functional split is under development on the basis of \gls{OAI} code (see \cite{fs6}).

\subsection{Runtime analysis}


Beyond the hard real-time constraints involved in the base band processing of radio signals, the major performance problem of Cloud-RAN is due to the non-deterministic behavior of the channel coding function. Much of this variability is due to radio channel conditions of \glspl{UE} attached to the base station, the data load per \gls{UE}, as well as the amount of traffic in the cell. The above observations raise fundamental questions with regard to conceiving virtual RAN functions and dimensioning the required computing capacity to execute them in the Cloud.

\begin{figure}[hbtp]
  \centering
   \includegraphics[scale=\tallafigurabigg, trim=0 40 0 0, clip] {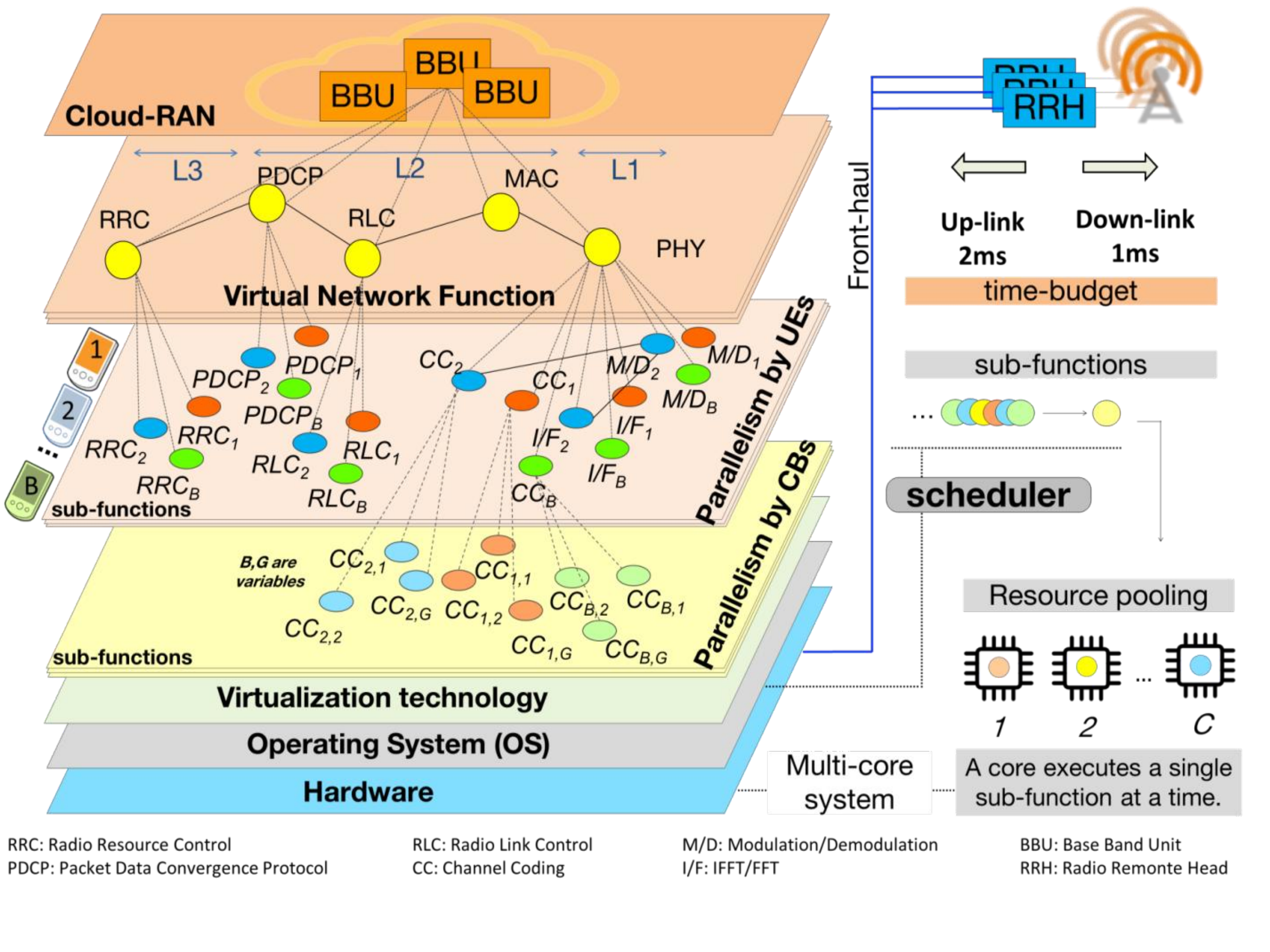}
    \caption{Multi-threading model.}
  \label{fig:cran_dataunits1}
\end{figure}

To achieve important gains in terms of latency, we have introduced a multi-threading model for executing the channel encoding/decoding function by means of massive parallel processing in a multi-core system. Enabling parallelism as much as possible and avoiding chaining when designing VNFs are fundamental principles to gain from the available computing resources. Chaining impacts are analyzed in~\cite{networks}. The proposed threading model uses the fact that radio sub-frames are composed of \gls{TB} whose content belongs to a single \gls{UE} for running UEs in parallel. In addition, when the size of a \gls{TB} is too big, it is segmented into shorter data units, referred to as \glspl{CB}. A \gls{CB} represents the smallest processing unit, which can be executed in parallel. As shown in  Figure~\ref{fig:cran_dataunits1}, \glspl{VNF} (e.g. vBBUs) appear as a chain of sub-functions which are executed on the top of the virtualization layer while sharing the available computing resources. A global scheduler is in charge of allocating the capacity of servers.


\section{Cloud-RAN modeling for dimensioning purposes}
\label{design}

\subsection{Job scheduling}

 
 
 We model parallel runnable jobs of VNFs as batches arriving at a processing facility composed of several cores. The objective then is to select the best scheduling strategy to meet latency requirements.  In this PhD thesis, we have considered several scheduling solutions (under greedy, round robin and dedicated principles) which can apply to a large range of virtualized network functions. Detailed results are not presented in this paper but can be found in~\cite{networks}. For Cloud-RAN, we have specifically considered two queuing models:
\begin{itemize}
    \item $M^{[X]}/M/C$ as an illustration of greedy disciplines (batches are queued and the head of line job is served as soon as a core is available);
    \item  $M^{[X]}/M/1$ Processor Sharing (PS) as a round robin strategy (all jobs are treated in parallel, there is no delay before accessing servers, but jobs receive an equal share of the global capacity).
\end{itemize} 
The two above strategies respectively illustrate parallel and concurrent computing. In parallel computing, each job runs on a single core and only one at any instant~\cite{pike2012concurrency}. Conversely, concurrent computing enables the simultaneous execution of jobs on a single core by overlapping time-periods; this leads to \gls{PS} models~\cite{kleinrock2,mxm1ps}. However, the drawback of processor sharing is in that multitasking on the same core requires context switching and memory splitting, which may considerably increase the latency. 

In the framework of the present PhD thesis, we have obtained new results for both models. The sojourn time 
 of a job in the $M^{[X]}/M/1$-PS queue has been obtained in \cite{mxm1ps}; we have acquired the probability survival function of the sojourn time, which has enabled us to derive the tail of the probability distribution. This result extends a previous result obtained by Kleinrock \emph{et al} in the 70's for the mean value \cite{kleinrock3}. The derivation of the sojourn time of a job is much more challenging as it involves correlations between the sojourn time of jobs. This issue is still under investigation.
 
 For the $M^{[X]}/M/C$ queuing system, we have obtained the Laplace transform of the sojourn time of a batch (on the basis of \cite{Cromie} by Cromie \emph{et al}) and derived the tail distribution.  We subsequently use this system to dimension the required amount of computing resources for the Cloud-RAN implementation, i.e., we use parallel processing in a strict sense so that jobs are simultaneously executed on separate cores avoiding latency introduced by time-sharing processors~\cite{rodriguez2017vnf,rodriguez2017towards}.

\subsection{Cloud-RAN modeling}

From a modeling point of view, each radio element, \gls{RRH}, belonging to a Cloud-RAN system represents a source of jobs in the up-link direction; while for the down-link direction, jobs arrive from the core network, which provides connection to external networks (e.g., Internet or other service platforms). There are then two queues of jobs for each radio element (antenna, DU), one in each direction. Since the time-budget for processing down-link sub-frames is half of that for up-link ones, they might be executed separately on dedicated processing units.  However, dedicating processors to each queue is not an efficient way of using limited resources.

For dimensioning purposes, we assume that virtual \glspl{BBU} (notably, virtual encoding/decoding functions) are invoked according to a Poisson process, i.e., inter-arrival times of runnable BBU functions are exponentially distributed. This reasonably captures the fact that in Cloud-RAN systems there is a sufficiently great number of antennas, which are not synchronized. In fact, \glspl{RRH} are at different distances of the BBU-pool, furthermore when considering no dedicated links, the fronthaul delay (inter-arrival time) can strongly vary because of network traffic. The occurrence of jobs then results from the superposition of independent point processes which justifies the Poisson assumption. The Poisson assumption is in some sense a worst case assumption with regard to fixed relative phases\footnote{In the same way as an $M/D/1$ queue is ``worse'' with regard to waiting time than an $\sum_i N_i D_i / D/1$ queue.}.

\begin{figure}[hbtp]
  \centering
   \includegraphics[scale=\tallafigura, trim=0 20 0 0, clip] {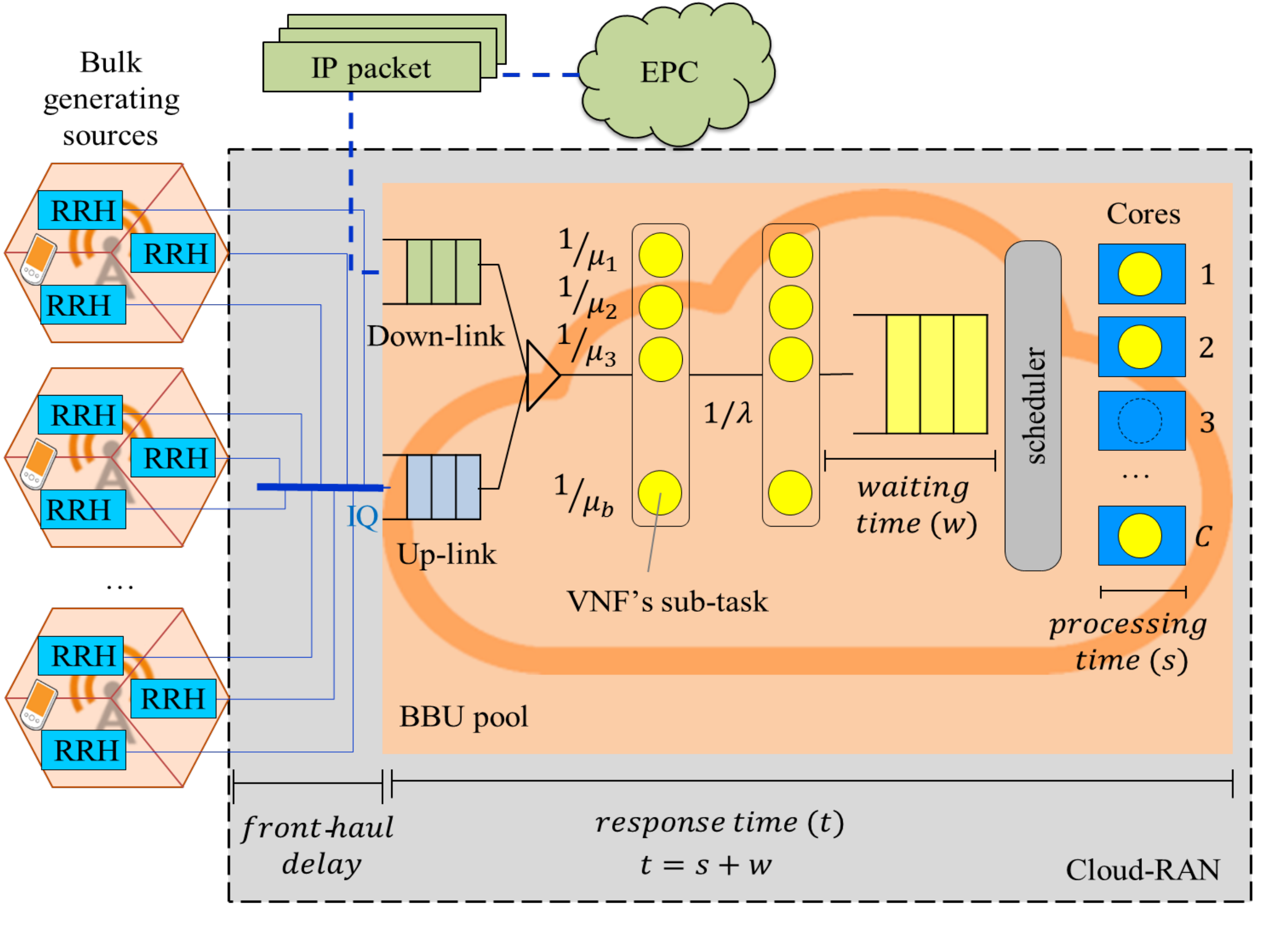}
    \caption{Stochastic service model for Cloud-RAN.}
  \label{fig:cran_model}
\end{figure}

The parallel execution of RAN functions  on a multi-core system with $C$ cores can then be modeled by bulk arrival systems, namely, an $M^{[X]}/G/C$ queuing system~\cite{valuetools2017}. As shown in Figure~\ref{fig:cran_model}, we consider each task-arrival to be in reality the arrival of $B$ parallel runnable sub-tasks or jobs, $B$ being a random variable. Each sub-task requires a single stage of service with a general time distribution. The runtime of each sub-task depends on the workload as well as on the network sub-function that it implements.
The number of parallel runnable sub-tasks belonging to a network sub-function is variable. Thus, we consider a non fixed-size bulk to arrive at each request arrival instant. The inter-arrival time is exponential with rate $\lambda$. The batch size $B$ is independent of the state of the system. When assuming that the computing platform has a non-limited buffer, the stability of the system requires $\rho={\lambda\E[B]}/{\mu C}< 1$.

We further assume that the processing time of a job is exponentially distributed with mean  $1/\mu$. This assumption is intended to capture the randomness in the runtime of \glspl{UE} due to the non-deterministic behavior of the channel coding function. When considering that the number $B$ of \glspl{UE} per subframe is geometrically distributed with mean $1/(1-q)$ (i.e., $\P(B=k) =(1-q)q^{k-1} $ for $ k\geq 1$), the complete service time of a radio subframe is then exponentially distributed with mean $1/((1-q)\mu)$. The geometric distribution as the discrete analog of the exponential distribution capture the variability of scheduled \glspl{UE} in a subframe. From the above analysis, the $M^{[X]}/M/C$ model enables the evaluation of the runtime of a subframe in a Cloud-RAN architecture based on parallel processing in a multi-core platform. The batch model as well as the analysis of the Markovian chain are presented in~\cite{jsac}.



\subsection{Cloud-RAN dimensioning}

The final goal of Cloud-RAN modeling is to determine the amount of computing resources needed in the cloud to guarantee the base-band processing of a given number of cells within deadlines. For this purpose, we evaluate the $M^{[X]}/M/C$ model while increasing $C$, until an acceptable probability of deadline exceedance (say, $\varepsilon$). The required number of cores is then the first value that achieves $P(T>\delta)<\varepsilon$, where $T$ represents the sojourn time of a batch, $\delta$ is a prescribed deadline and $\varepsilon$ is a tolerance for sojourn time exceedance and hence for the loss probability of an entire batch. 
 
\begin{figure}[hbtp]
  \centering
   \includegraphics[scale=\tallafigura, trim=60 20 0 0, clip] {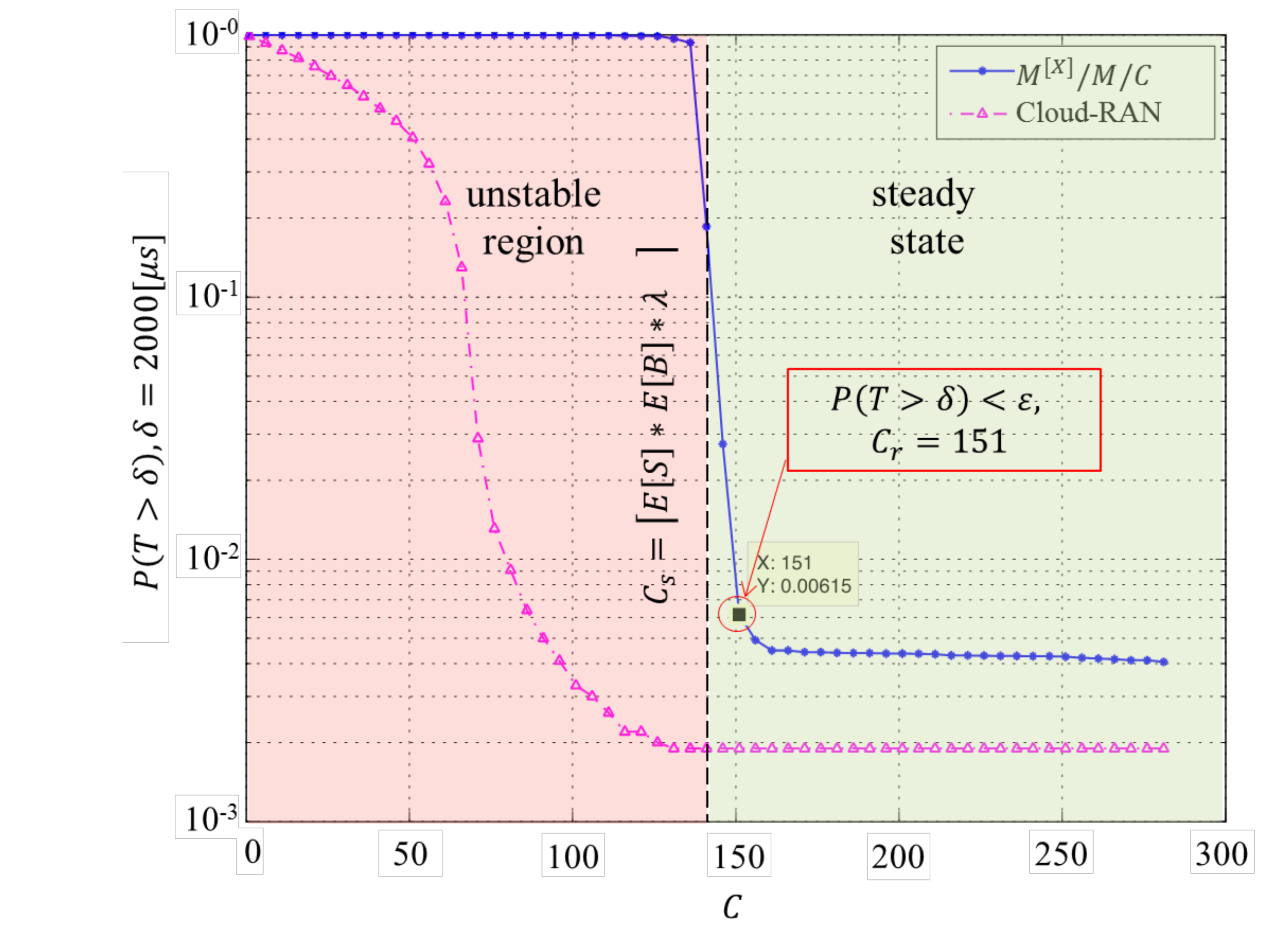}
    \caption{Cloud-RAN dimensioning using the $M^{[X]}/M/C$ model.}
  \label{fig:cran_dataunits}
\end{figure}

 We validate by simulation the effectiveness of the $M^{[X]}/M/C$ model with the behavior of the real Cloud-RAN system hosting $100$ \glspl{eNB} of $20$ MHz during the reception process. As illustrated in Figure~\ref{fig:cran_dataunits}, we observe that for a given $\varepsilon=0.00615$, the required number of cores is $C_{r}=151$, which is in accordance with the real C-RAN performance, where the probability of deadline exceedance is barely $0.00018$.  When $C$ takes values lower than a certain threshold $C_{s}$, the C-RAN system is overloaded, i.e., the number of cores is not sufficient to process the vBBUs workload; the system is then unstable.

 %



We have additionally studied the impatience criterion (which reflects RAN time-budgets). Results showed that impatience is not incident when the probability of deadline exceedance is low enough. See~\cite{jsac} for details. A worst-case analysis regarding the dimensioning problem was depicted in~\cite{rodriguez2017performance}.


\section{Proof of Concept}
\label{poc}
\subsection{Testbed description}

As a \textit{proof of concept}, we have implemented on the basis of various open-source solutions notably \gls{OAI}, an end-to-end virtualized mobile network. This platform notably implements the proposed models and scheduling strategies. The parallel processing of both encoding (downlink) and decoding (uplink) functions is carried out by using multi-threading in a multi-core server within a single process. Latency is considerably reduced since we avoid multi-tasking across different processes. The workload of threads is managed by a global non-preemptive scheduler. Each thread is assigned to a dedicated single core with real-time OS priority and is executed until completion without interruption. The isolation of threads is provided by a specific configuration performed in the OS which prevents from the use of channel coding computing resources for any other job.

\begin{figure}[hbtp]
  \centering
   \includegraphics[scale=\tallafigurabig, trim=138 0 0 0, clip] {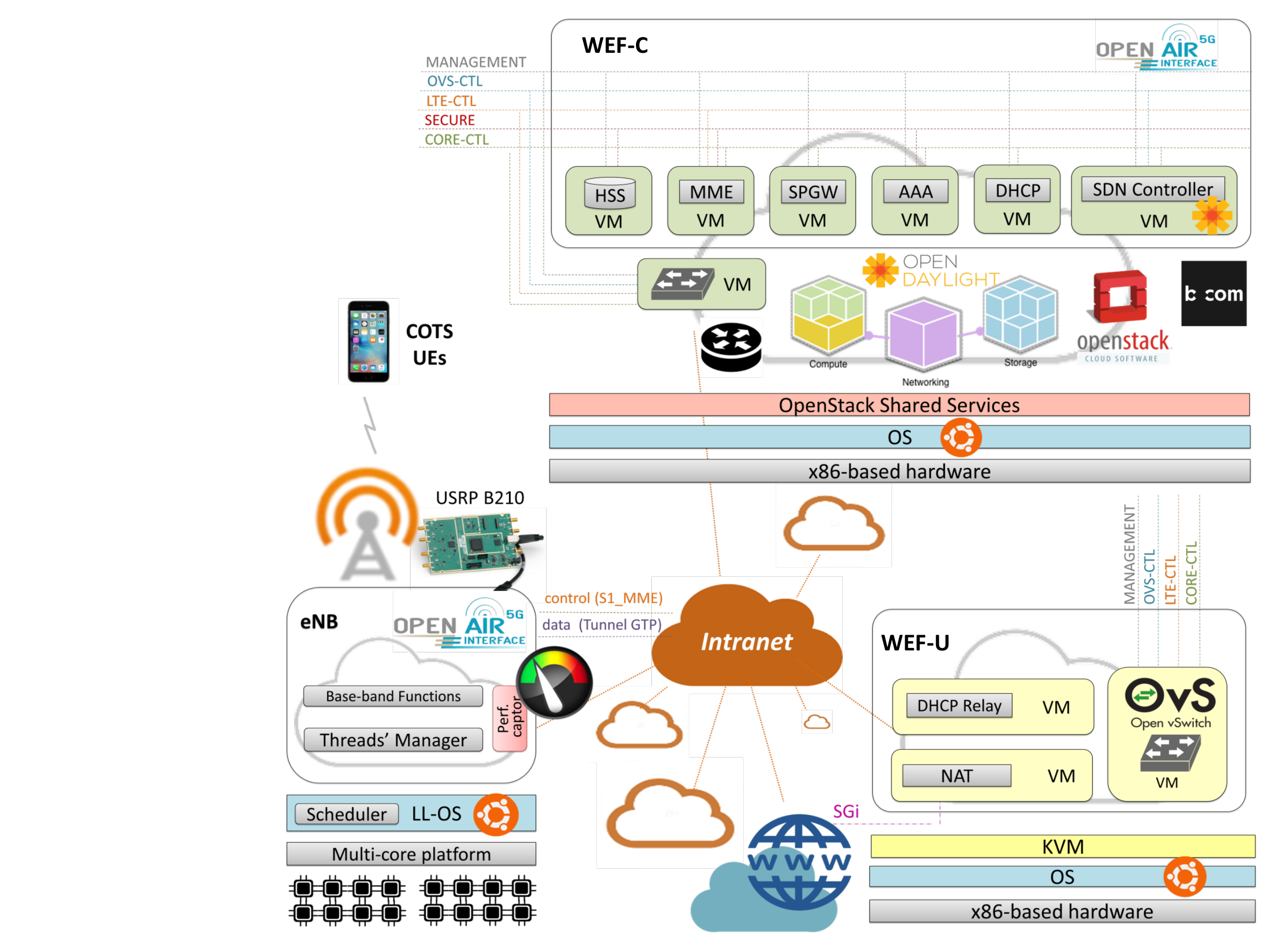}
    \caption{Proof of Concept, multi-threading implementation.}
  \label{fig:PoC}
\end{figure}

As shown in Figure~\ref{fig:PoC},  the platform is based on a complete separation of the user and control plane as recommended by 3GPP for 5G networks, referred to as \gls{CUPS}, and implemented in b$<>$com’s solution (Wireless  Edge Factory, WEF). As virtualization engines we use KVM, OpenStack, and OpenDaylight. The radio element of the mobile network is performed by an USRP B210 card. Commercial smartphones can be then connected. 


When a UE attaches to the network, the AAA (Authentication, Authorization, and Accounting) procedure is triggered by the MME; user profiles are validated by the HSS. When access is granted to the UE, the DHCP component provides it the IP-address. The end-to-end connection is assured after the creation of the GTP-U and GTP-C tunnels. The NAT component provides address translation and is deployed between the SGi interface and the Internet network.

\subsection{Performance Results}

The behavior of Cloud-RAN in terms of latency when performing parallel processing has been evaluated by both simulation and testbed. The simulation considers a Cloud-RAN system of $100$ eNBs running in a data center equipped with $151$ cores (acquired with the $M^{[X]}/M/C$ model). Results lead relevant conclusions in terms of scalability.

\begin{figure}[hbtp]
  \centering
   \includegraphics[scale=\tallafigura, trim=40 20 0 0, clip] {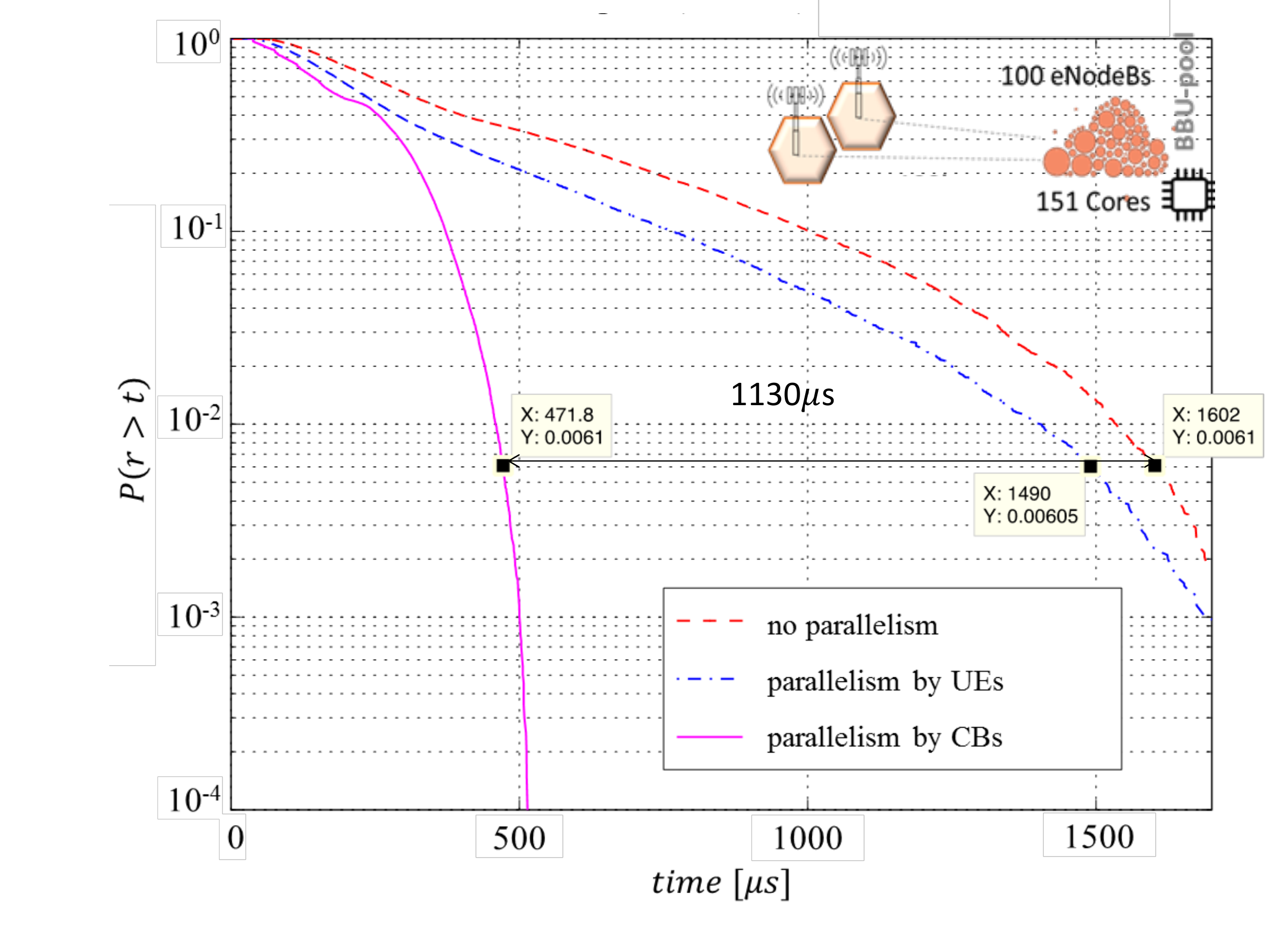}
    \caption{Cloud-RAN performance while considering $100$ eNBs.}
  \label{fig:simgain}
\end{figure}

Figure~\ref{fig:simgain} shows the CDF of the sojourn time of radio sub-frames when performing parallel programming. It is observed that more than $99$\% of sub-frames are processed within $472$ microseconds and $1490$ microseconds when performing parallelism by \glspl{CB} and \glspl{UE}, respectively. It represents a gain of $1130$ microseconds (\gls{CB}) and $100$ microseconds (\gls{UE}) with respect to the original system (non-parallelism). These gains in the sojourn time enable network operators to increase the maximum distance between antennas and the central office. Hence, when considering the light-speed in the optic-fiber, i.e., $2.25*10^8$m/s, the distance can be increased up to $\approx250$ km when running \glspl{CB} in parallel. 

\begin{figure}[hbtp]
  \centering
   \includegraphics[scale=\tallafigura, trim=40 0 0 0, clip] {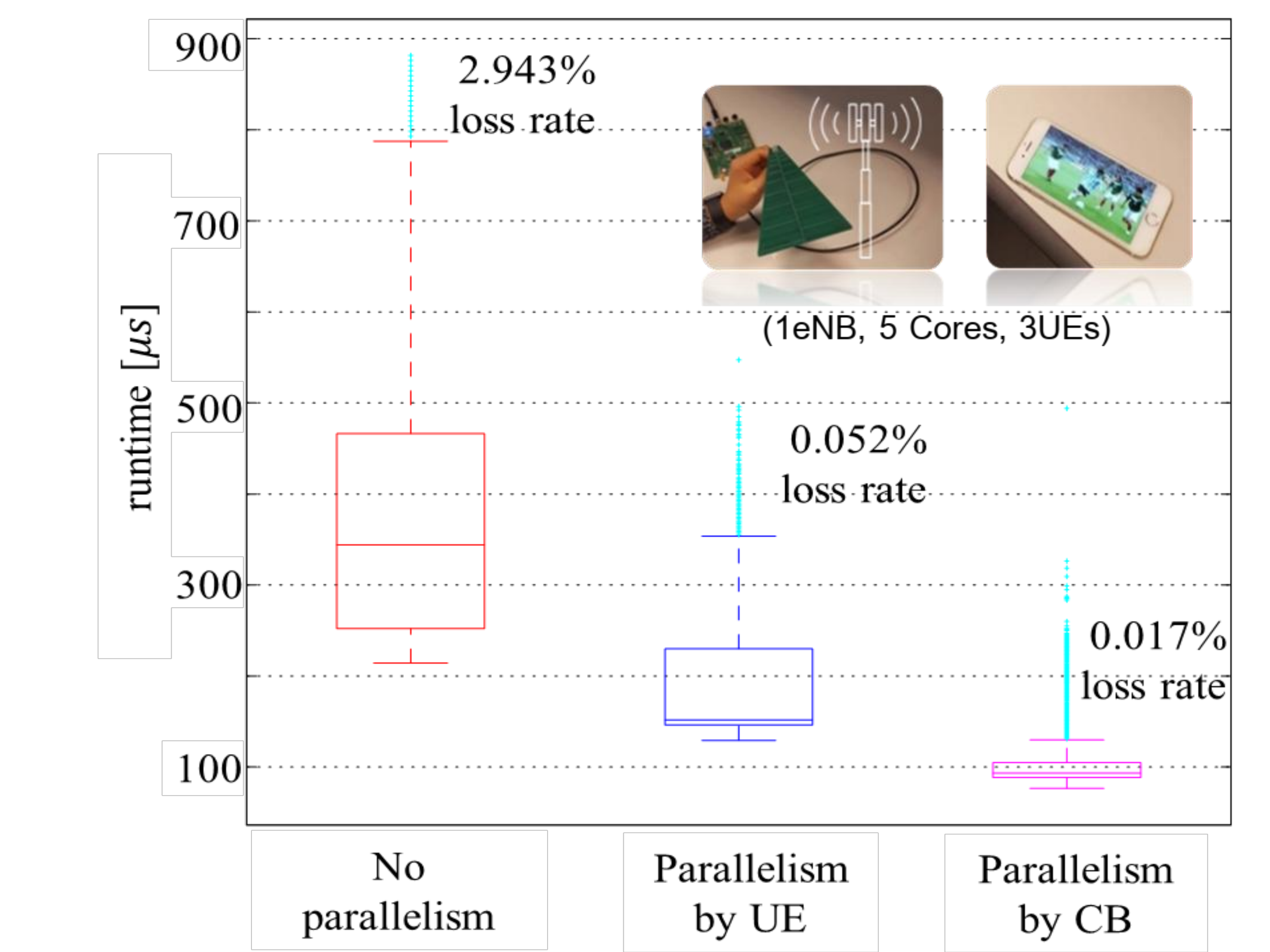}
    \caption{Testbed performance results (Uplink).}
  \label{fig:perfresults}
\end{figure}

The testbed implementation considers a single virtualized \gls{BBU} (namely, an \gls{eNB} of $20$ MHz) and three \glspl{UE}. As shown in Figure ~\ref{fig:perfresults}, the testbed confirmed the accuracy of the performance gains obtained by simulation~\cite{jsac}. Decoding function shows a performance gain of $72,6\%$ when executing CBs in parallel. Results open the door for deploying fully centralized cloud-native RAN architectures.

\section{Conclusion and Research perspectives}
\label{conclusion}

We have introduced in this PhD thesis a solution for implementing and dimensioning Cloud-RAN systems. The solution is based on parallel processing of the channel coding function in multi-core platforms. This principle significantly reduces latency of RAN functions and thus enables the distance between \gls{DU} and \gls{CU} to be significantly increased (achieving as a consequence a higher aggregation of eNBs in the cloud for taking advantage of resource pooling and statistical multiplexing principles). The proposed solution has been implemented on a testbed on the basis of OAI open source RAN code.

Furthermore, we have investigated and modeled two scheduling strategies which have led us to establish new results for two queuing systems (namely, the $M^{[X]}/M/C$ and $M^{[X]}/M/1$-PS), both widely studied in the queuing literature. The $M^{[X]}/M/C$ can then be used to dimension the multi-core platform supporting the Cloud-RAN. 

Beyond increased distance between \gls{CU} and \gls{DU}, we have designed an intra-PHY functional split (Figure~\ref{figsplit}) which enables the reduction of the required bandwidth in the fronthaul network. This split has been implemented on OAI code and exhibits excellent performance results in terms of bandwidth \cite{fs6}. Since the bit rate between \gls{DU} and \gls{CU} is no more constant (in contrary to classical CPRI) it is envisaged to develop statistical multiplexing strategies between the radio and the optical layers in  order to optimize the use of the bandwidth in the fronthaul~\cite{PatentNejm}.

One major outcome of this PhD thesis is to show that critical RAN functions can be virtualized and then orchestrated as any other network function. In \cite{NoF}, we have shown how a complete mobile network can be orchestrated with a carrier grade automation platform, namely the Open Network Automation Platform (ONAP), in the context of network slicing. 


\bibliographystyle{IEEEtran}
\bibliography{biblo}

\pagebreak
\end{document}